\documentclass{pasj00}

\newcommand{\gsim}{\raisebox{0.3mm}{\em $\, >$} \hspace{-3.3mm}
\raisebox{-1.8mm}{\em $\sim \,$}}
\newcommand{\bm}[1]{\mbox{\boldmath $#1$}}

\begin{document}
\SetRunningHead{Ohsuga et al.}
{RMHD Simulations of Accretion Flow and Outflow}
\Received{0000/00/00}
\Accepted{0000/00/00}

\title{Global Radiation-Magnetohydrodynamic Simulations 
of Black Hole Accretion Flow and Outflow: 
Unified Model of Three States}

\author{Ken \textsc{Ohsuga},\altaffilmark{1,2}
        Shin \textsc{Mineshige},\altaffilmark{3}
        Masao \textsc{Mori},\altaffilmark{4}
        and
        Yoshiaki \textsc{Kato}\altaffilmark{5}
        }
\altaffiltext{1}{National Astronomical Observatory of Japan, Osawa, Mitaka, Tokyo 181-8588}
\altaffiltext{2}{Institute of Physical and Chemical Research (RIKEN), Hirosawa, Wako, Staitama 351-0198}
\altaffiltext{3}{Department of Astronomy, Graduate School of Science, Kyoto University, Kyoto 606-8502}
\altaffiltext{4}{Center for Computational Sciences, University of Tsukuba, Tsukuba, Ibaraki 305-8577}
\altaffiltext{5}{Institute of Institute of Space and Astronautical
        Science (JAXA),
Yoshinodai, Sagamihara, Kanagawa 229-8510}


\maketitle


\begin{abstract}
Black-hole accretion systems are known to 
possess several distinct modes (or spectral states),
such as low/hard state, high/soft state, and so on.
Since the dynamics of the corresponding flows is distinct,
theoretical models were 
separately discussed for each state. 
We here propose a unified model based on our new, global, 
two-dimensional radiation-magnetohydrodynamic simulations.
By controlling a density normalization we could for the first time 
reproduce three distinct modes of accretion flow and outflow
with one numerical code.  
When the density is large (model A), 
a geometrically thick, very luminous disk forms,
in which photon trapping takes place.
When the density is moderate (model B), 
the accreting gas can effectively cool 
by emitting radiation, thus generating a thin disk,
i.e., the soft-state disk. 
When the density is too low for radiative cooling to be important
(model C), 
a disk becomes hot, thick, and faint; i.e., the hard-state disk.
The magnetic energy is amplified within the disk up to about twice, 
30\%, and 20\% of the gas energy 
in models A, B, and C, respectively. 
Notably, the disk outflows with helical magnetic fields,
which are driven either by radiation pressure force
or magnetic pressure force,
are ubiquitous in any accretion modes.
Finally, our simulations are consistent with
the phenomenological $\alpha$-viscosity prescription,
that is, the disk viscosity is proportional to the pressure. 
\end{abstract}



\section{Introduction}
%
The extensive study of disk accretion flows
started in the 1960's. 
The standard disk model, and then 
the slim disk model  
and the radiatively inefficient accretion flow (RIAF) model 
were 
proposed for explaining a variety of accretion modes
(\cite{SS73}; \cite{Ichi77}; 
\cite{Rees82}; \cite{Abramo88};
\cite{NY94}). 
These models are successful, but have some limitations.
For example, the disk viscosity, the most important key ingredient 
for the accretion disk theory, 
is prescribed by a phenomenological $\alpha$-viscosity model,
whereby the viscous torque is proportional to the pressure 
\citep{KFM08}, 
although its physical basis is not clear. 
They are (radially) one-dimensional models 
so that they cannot describe multi-dimensional motion, 
such as outflow and internal circulation. 
Complex coupling between radiation, magnetic fields, 
and matters is not accurately solved, either. 

Since the disk viscosity 
is likely to be of magnetic origin \citep{BH91},
multi-dimensional global magneto-hydrodynamics (MHD) 
simulations are being rather extensively performed recently 
as a model for the disks with low luminosities 
(\cite{Matsumoto99}; \cite{MHM00}; \cite{HK01};
\cite{Koide01}; \cite{DV03}; \cite{HK06})
. 
%
%
Such non-radiative MHD simulations 
cannot explain higher luminosity states, however, 
since strong matter-radiation coupling is expected.  
As an independent approach several groups performed 
two-dimensional radiation-hydrodynamic (RHD) simulations 
of very luminous flow since the 1980's 
(\cite{ECK88}; \cite{Okuda00}; \cite{O05}; \cite{O06}).  
Those simulations were, however, non-MHD simulations 
and so they were obliged to rely on the phenomenological 
$\alpha$-viscosity model.  
%
Multi-dimensional radiation-MHD (RMHD) simulations
are unavoidable.
Such simulations were attempted in the past
(e.g., \cite{Turner03}; \cite{Hirose06}),
but these are restricted to local simulations performed
under the shearing-box approximations
and, hence, global coupling of magnetic fields was
artificially quenched there.

We, here, report for the first time the results of 
global two-dimensional RMHD simulations
with a motivation to establish a unified view of
the accretion flow and outflow around the black holes.  

\section{Numerical Method}
Our method of calculations is extension of 
that of MHD simulations (e.g., \cite{KMS04}).
We use cylindrical coordinates ($r$, $\varphi$, $z$), 
where $r$ is the radial distance, 
$\varphi$ is the azimuthal angle, 
and $z$ is the vertical distance. 
We assume that the flow is non-self-gravitating, 
reflection symmetric relative to equatorial plane, 
and axisymmetric with respect to the rotation axis. 
General relativistic
effects are incorporated by the pseudo-Newtonian potential 
\citep{PW80}. 
For the opacity, we consider the Thomson scattering,
free-free absorption, and bound-free absorption 
(\cite{RL79}; \cite{HHS62}). 
The energy equations of gas and radiation are given by
\begin{eqnarray}
 \frac{\partial E_{\rm gas}}{\partial t}
  &+& \nabla\cdot(E_{\rm gas} {\bm v}) \nonumber\\
  &=& -p_{\rm gas}\nabla\cdot{\bm v} -4\pi \kappa B 
  + c\kappa  E_{\rm rad}
  + \frac{4\pi}{c^2}\eta J^2,
  \label{gase}
\end{eqnarray}
and
\begin{eqnarray}
 \frac{\partial E_{\rm rad}}{\partial t}
  &+& \nabla\cdot(E_{\rm rad} {\bm v}) \nonumber\\
  &=& -\nabla\cdot{\bm F_{\rm rad}} -\nabla{\bm v}:{\bm {\rm P}_{\rm rad}}
  + 4\pi \kappa B - c\kappa E_{\rm rad},
  \label{rade}
\end{eqnarray}
where 
$E_{\rm gas}$ is the internal energy density of the gas,
$\bm{v}$ is the velocity, 
$p_{\rm gas}$ is the gas pressure ($\equiv 2E_{\rm gas}/3$),
$B$ is the blackbody intensity,
$J$ is the electric current,
$E_{\rm rad}$ is the radiation energy density,
${\bm F}_{\rm rad}$ is the radiative flux,
${\bm {\rm P}}_{\rm rad}$ is the radiation pressure tensor,
$\kappa$ is the absorption opacity.
We adopt the anomalous resistivity, $\eta$, 
which is the same as that used in \citet{KMS04};
\begin{equation}
 \eta=\left\{
       \begin{array}{ll}
	0 & \mbox{for } v_{\rm d}<v_{\rm crit} \\
	\eta_{\rm max}
	 \left[\left(\frac{v_{\rm d}}{v_{\rm crit}} \right)-1 \right]^2 
	 & \mbox{for } v_{\rm crit}\leq v_{\rm d}<2v_{\rm crit} \\
	\eta_{\rm max} & \mbox{for } v_{\rm d}\geq 2v_{\rm crit} 
       \end{array}
       \right.
 ,
\end{equation}
where 
$v_{\rm d}\equiv J/\rho$ 
is the electron drift velocity,
$v_{\rm crit}\equiv 0.01c$ 
is the critical velocity,
and $\eta_{\rm max}\equiv 10^{-3}cR_S$
is the maximum resistivity
with $R_S$ being the Schwarzschild radius. 
This form of the anomalous resistivity was proposed by 
\citet{Yokoyama94} to account for the occurrence 
of fast reconnections in solar flares.
The MHD related terms are solved by the modified Lax-Wendroff scheme 
\citep{RB67}.
We employ the flux-limited diffusion (FLD) approximation 
to solve the radiation energy equation \citep{LP81}.
The radiation energy transport via the radiative flux 
is solved based on the implicit method, where 
we separately treat radiative fluxes in the radial and vertical 
directions with using Thomas method for a matrix inversion.
The gas-radiation interaction is also
solved with the implicit method, 
which is basically the same as 
that described by \citet{TS01}.
An advection term in the energy equation of the radiation
is solved with the explicit method,
in which an integral formulation is used to generate 
a conservative differencing scheme.
We performed the test of two-dimensional radiation propagation \citep{TS01},
finding that the energy loss is less than 0.005\% in 600 steps, from
which we estimate the error in the energy conservation not to exceed 0.08\%
during the photon traveling timescale in our simulations.

The grid extends from $3R_S$ to $103R_S$ in the radial direction 
and from $0$ to $91.6R_S$ in the vertical direction.
The grid spacing is uniform, $0.2R_S$, in both directions. 
We adopt free boundary conditions for the matter and magnetic fields; 
i.e., the matter can freely go out but not to come in and
the magnetic fileds do not change across the boundary. 
We assume that radiation goes out with the radiation flux of 
$c E_{\rm rad}$,
except at $r=3R_S$ and $z>3R_S$, 
through which no radiation goes out. 
We assume a black hole mass to be $10M_\odot$. 
In future we will perform simulations, in which we will put the boundary 
condition at around $R_S$ and to solve the both side of the midplane.

We start calculations with a rotating torus,
in which the magnetic fileds are purely poloidal
(plasma-$\beta=100$) and closed loops in the torus, 
being located at around $40R_S$ embedded in non-rotating isothermal corona. 
Our initial conditions are the same as those of model B in Kato et al. (2004), 
except that the density of the corona is 0.05 times as large as that one.
We evolve the initial torus by solving non-radiative MHD equations 
for $1$ sec.
We then 
assign the density normalization ($\rho_0$), 
density at the center of the initial torus,
and turn on the radiation terms. 
We calculate three models in total, by setting 
$\rho_0=1 \rm g\,cm^{-3}$ (model A), 
$10^{-4} \rm g\,cm^{-3}$ (model B), 
and $10^{-8} \rm g\,cm^{-3}$ (model C).  
Since radiation loss rate depends on the density, 
we can reproduce the three distinct regimes of accretion flow.

\section{RESULTS}
\subsection{Accretion Flows}
\begin{figure*}[t]
  \begin{center}
    \FigureFile(180mm,180mm){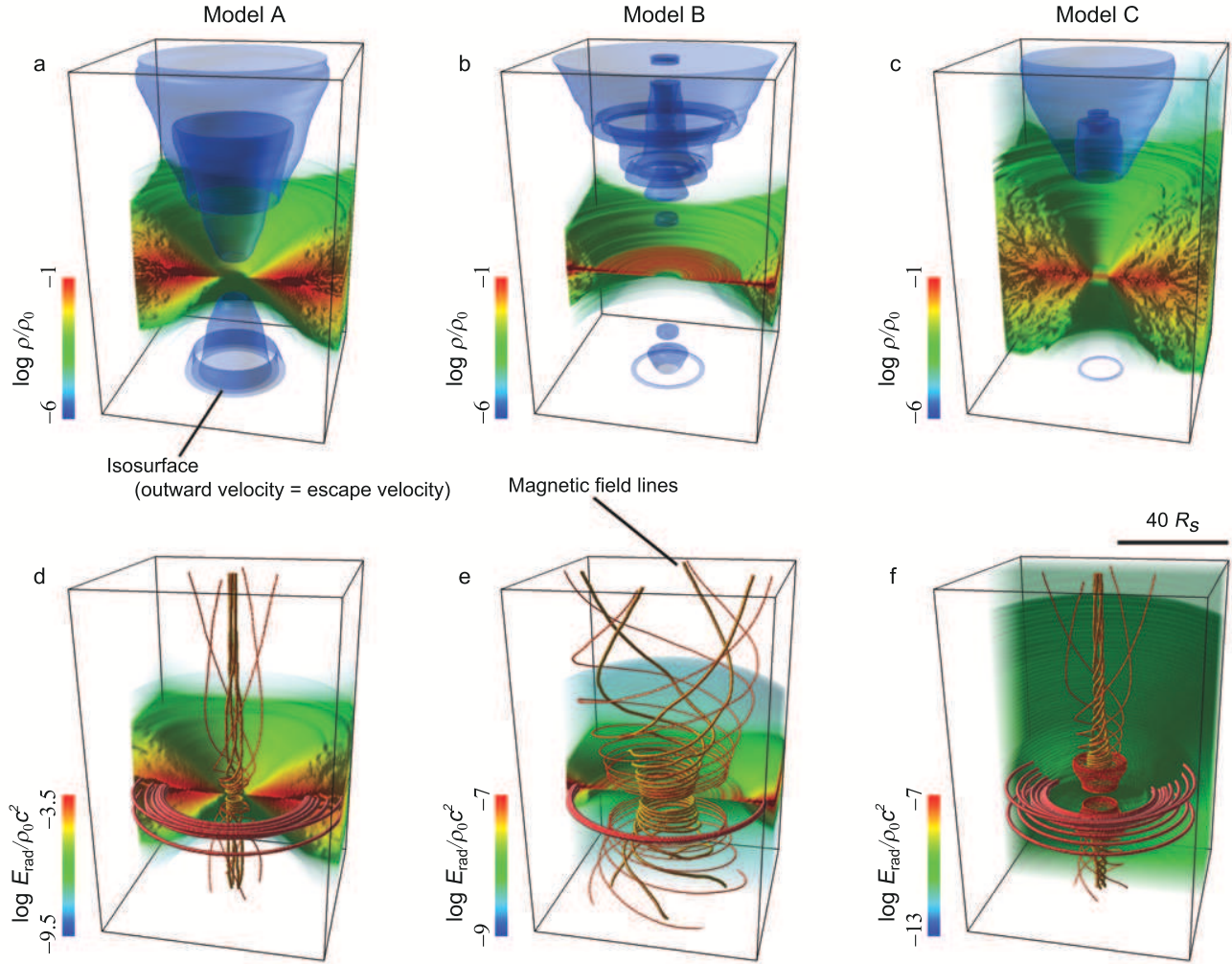}
  \end{center}
 \caption{
Perspective view of inflow and outflow patterns near the black hole
for models A, B, and C, from left to right, respectively. 
Upper panels:
Normalized density distributions (color) are overlaid with isosurfaces, 
at which the outward velocity equals to the escape velocity.
Here, the elapse times are 5.25 sec, 7.75 sec, and 5.0 sec
for models A, B and C, respectively. 
Lower panels:
The distributions of the normalized radiation energy density
(color) is overlaid with the magnetic field lines.
In models A and B, the green surfaces of the disks 
correspond to the photosphere, 
where the optical thickness measured from the upper 
($z=91.6R_S$) or lower ($z=-91.6R_S$) 
boundaries is unity. 
Note that the disk in model C is optically thin. 
}
 \label{fig1}
\end{figure*}

Figure \ref{fig1} clearly visualizes that the
flow patterns differ significantly among three models.
The typical mass accretion rate ($\dot{M}_{\rm acc}$),
the luminosity ($L$), and the density ($\rho$) and
the temperature ($T$), 
as well as other important quantities, 
are summarized in Table \ref{tab1}.
\begin{table*}
 \begin{center}
  \caption{
Three calculated models and basic properties.
}
  \begin{tabular}{llccc}
   \hline
   \multicolumn{2}{c}{
   \begin{tabular}{c}
      Model\\Corresponding States
   \end{tabular}}
   & \begin{tabular}{c} 
      A\\Slim disk
     \end{tabular}
   & \begin{tabular}{c} 
      B\\Standard disk
     \end{tabular}
   & \begin{tabular}{c} 
      C\\RIAF
     \end{tabular} \\
   \hline
   normalized mass accretion rate\footnotemark[$*$] & $\dot{M}_{\rm acc}/(L_{\rm E}/c^2)$  & $120$ & $5.3\times 10^{-3}$ & $5.8\times 10^{-5}$ \\
   Eddington ratio\footnotemark[$*$] & $L/L_{\rm E}$ & $1.0$ & $7.3\times 10^{-4}$ & $7.4\times 10^{-10}$ \\
   ratio of outflow rate to accretion rate\footnotemark[$*$] & $\dot{M}_{\rm out}/\dot{M}_{\rm acc}$ & 0.045 & 0.013 & 0.1 \\
   ratio of kinetic luminosity to luminosity\footnotemark[$*$] & $L_{\rm kin}/L$ & 0.16 & 0.0013 & 120 \\
   density\footnotemark[$\dagger$] & $\log \rho \,[\rm g\,cm^{-3}]$ & $-1\sim -2$ & $\sim -5$ & $\sim -10$ \\
   gas temperature\footnotemark[$\dagger$] & $\log T \,[K]$ & $\sim 8$ & $\sim 6$ & $10\sim 11$ \\
   gas energy density\footnotemark[$\ddagger$] & $\log E_{\rm gas} \,[\rm erg\,cm^{-3}]$ & 15.0 & 9.7 & 9.0 \\
   magnetic energy density\footnotemark[$\ddagger$] & $\log E_{\rm mag} \,[\rm erg\,cm^{-3}]$ & 15.3 & 9.2 & 8.3 \\
   radiation energy density\footnotemark[$\ddagger$] & $\log E_{\rm rad} \,[\rm erg\,cm^{-3}]$ & 17.2 & 10.4 & 3.2 \\
   plasma-$\beta$\footnotemark[$\ddagger$] & $p_{\rm gas}/p_{\rm mag}$ & 0.33 & 2.1 & 3.3 \\
   \hline
   \multicolumn{4}{@{}l@{}}{\hbox to 0pt{\parbox{170mm}{\footnotesize
   \par\noindent
   \footnotemark[$*$] 
   The mass accretion rate ($\dot{M}_{\rm acc}$) 
   is evaluated the mass passing through 
   the inner boundary near the black hole ($r=3R_S$ and $z<3R_S$)
   per unit time,
   and outflow rates ($\dot{M}_{\rm out}$) indicate
   the mass passing through the upper boundary ($z=91.6R_S$)
   with higher velocities than the escape velocity
   (high-velocity outflow) per unit time.
   The luminosity ($L$) is calculated 
   based on the radiative flux at the upper boundary,
   and the kinetic luminosity ($L_{\rm kin}$) is 
   calculated from the amount of the kinetic energy carried 
   by the high-velocity outflow.
   These values are time-averaged over $t=5.5-6.0$ sec (models A and C) 
   or $t=7.5-8.0$ sec (model B). 
   \par\noindent
   \footnotemark[$\dagger$] Time-averaged over $t=5.0-6.0$ sec 
   (models A and C) or $t=7.0-8.0$ sec (model B) 
   at the regions $r=10-20R_S$ and $z\sim 0$. 
   \par\noindent
   \footnotemark[$\ddagger$] Density-weighted spatial averages 
   within the regions of $r=10-20R_S$ and $z=0-10R_S$ and 
   are time-averaged over $t=5.0-6.0$ sec (models A and C) 
   or $t=7.0-8.0$ sec (model B). Here, $p_{\rm gas}$ and $p_{\rm mag}$
   are gas and magnetic pressures.
   }\hss}}
  \end{tabular}
  \label{tab1}
 \end{center}
\end{table*}

In model A with a relatively large density normalization, 
the mass accretion rate exceeds the Eddington rate, $L_{\rm E}/c^2$, 
with $L_{\rm E}$ being the Eddington luminosity. 
The disk is optically and geometrically thick.
Photons are not easy to go out from the surface 
due to a large optical depth 
so that radiative cooling is restricted. 
We confirm the photon-trapping effects.
The disk is supported by radiation pressure. 
Circular motion appears in the disk region. 
Since $L\gsim L_{\rm E}$, 
this model corresponds to the two-dimensional version 
of the slim disk model. 
The calculated temperature and density are also consistent.

In model B with a moderate density normalization, 
a geometrically thin disk forms 
because of efficient radiative cooling. 
The disk is optically thick and supported 
mainly by radiation pressure, 
which is slightly greater than the gas pressures. 
Such properties, as well as the temperature and density, 
agree with those of the standard disk model.
It might be noted well that the flow in model B has not reached 
a quasi-steady state, 
since the viscous timescale is about 45 sec at 
$r=10R_S$, 
whereas the elapsed time is 8 sec.
In a forthcoming paper we will present finer mesh calculations
with adequate grid spacing, by which
we will be able to investigate the detailed, internal structure 
of the thin disk.


In model C with a small density normalization, 
the density is too low for radiative cooling 
to be important. 
The disk is filled with hot rarefied plasmas
and is geometrically thick but optically thin.
We find significant circular motion inside the disk. 
This model corresponds to the RIAF model.

\subsection{Outflows}
As shown in Figure \ref{fig1},
the disk outflows with helical magnetic fields
are ubiquitous around the black holes.
In models A and C
the magnetic field lines stretch out vertically 
in the vicinity of the rotation axis.
While, around the equatorial plane,
the toroidal component of the magnetic fields
is dominant over other components in all models,
which are reminiscent of magnetic-tower jets
(\cite{Lynden96}; \cite{KMS04}).

In model A, 
the strong radiation pressure force is responsible for 
driving the quasi-steady outflows above and below the disk, 
whose velocity amounts to $\sim 0.25c$. 
We find that the radiation energy density ($E_{\rm rad}$) is very large 
in the disk region, 
and the steep profile of $E_{\rm rad}$ enhances 
the radiation force (radiative flux) \citep{O05}. 

Remarkably, 
our simulation of model B shows 
the occurrence of magnetically powered disk wind, 
on the contrary to the usual belief regarding the standard disk model.
Note, however, that the disk wind is not so strong in model B;
$L_{\rm kin}\ll L$, 
and $L_{\rm kin}/L$ is the smallest among all models,
where $L_{\rm kin}$ is kinetic luminosity. 
It must be stressed that we have solved the entire 
inflow-outflow structure simultaneously 
unlike the previous simulations, 
in which the inflow (disk) structure was not solved 
but treated as the boundary condition 
\citep{Proga04}. 

Our RMHD simulations reveal $L_{\rm kin}/L>1$ in model C 
in contrast with models A and B, 
implying that the disks with $\dot{M}_{\rm acc}\ll L_{\rm E}/c^2$ 
lose the energy via the jets rather than via radiation.
The outflow rate is 10 \% of the mass accretion rate, 
and the ratio is largest in three models. 
The photons freely escape from the disk, 
producing quasi-spherical distribution of $E_{\rm rad}$, 
whereas the radiation energy is enhanced inside the disk 
in models A and B.
The radiation force is negligible 
because of small radiative flux.

The FLD approximation is good for Models A and C, since
the whole region (except for the very vicinity of the inner boundary) is 
optically thick for Thomson scattering in the former and
radiative cooling is never important in the latter.  
It is
known to be problematic to determine the direction of the radiation 
flux in regions where the optical depth is around unity (e.g., around the 
disk surface in model B). We, however, wish to stress that the outflow 
is accelerated by the magnetic pressure, and not by radiation force.

\subsection{Amplification of magnetic fields and viscosity}
We find that magnetic energy 
($E_{\rm mag}$) is amplified to be 30\% and 20 \% 
of the gas energy ($E_{\rm gas}$) in models B and C,
respectively (see Table \ref{tab1}). 
In model A, surprisingly, $E_{\rm mag}$ does exceed the gas energy,
$E_{\rm mag} \sim 2 E_{\rm gas}$.
Its implication is enormous: 
the viscosity had better been scaled in terms of the total 
(or radiation) pressure and not of the gas pressure 
in radiation pressure-supported disks
(cf., \cite{Sakimoto81}).

\begin{figure}
  \begin{center}
    \FigureFile(80mm,90mm){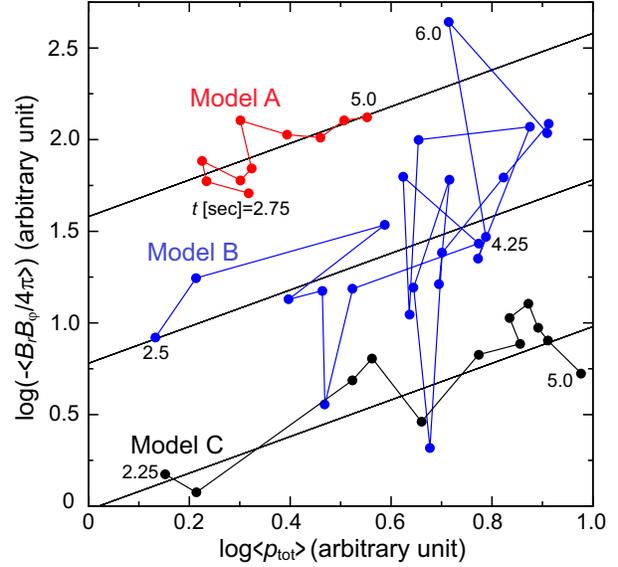}
  \end{center}
  \caption{
The density-weighted spatial averages of the magnetic torque near the
black hole, $r=5R_S$, 
as functions of the total pressure.
Each point represents the time-averaged value 
for every 0.25 sec. 
At the initial points,
($\log \left[-\langle B_r B_\varphi /4\pi \rangle\right]$, 
$\log\langle p_{\rm tot} \rangle$) $\sim$
($15.3$, $12.7$), ($10.9$, $8.7$), and ($8.6$, $5.8$)
in CGS unit for models A, B, and C, respectively.
}\label{fig3}
\end{figure}
It is one of the most significant issues in astrophysics 
how to prescribe the disk viscosity. 
In Figure \ref{fig3} we show how the magnetic torque, 
$\langle -B_r B_\varphi /4\pi \rangle$,
behaves as a function of the pressure in the region close to the
black hole (at $r=5R_S$).
We can see that the torque is roughly proportional 
to the total pressure (with some scatters) in all the cases. 
If we define the viscosity parameter as 
$\alpha=\langle -B_r B_\varphi/4\pi \rangle / \langle p_{\rm tot} \rangle$
where total pressure ($p_{\rm tot}$) is the sum of gas pressure 
and radiation energy density divided by 3, 
we estimate $\alpha\sim 0.004$, $0.006$, and $0.002$ 
for models A, B, and C, respectively.

Cautions should be taken to the point that
the properties of magnetic fields in the two-dimensional simulations 
could be affected by artificial occurrence of channel modes of flow 
and the anti-dynamo theorem and thus deviate from those of the 
three-dimensional simulations. Hence, the three-dimensional study 
should be explored in future work.

\section{Observational Implications}
Outflows and jets are ubiquitous in any regimes
of black-hole disk accretion flow.
They seem to manifest themselves by warm absorptions features 
of active galactic nuclei 
and by blue-shifted absorption lines in the black hole binaries (BHBs)
(\cite{Blustin05}; \cite{Cappi06}; \cite{Kubota07}).
The high-velocity outflows with velocity of $\sim 0.25c$ in model A 
will explain the X-ray observations
of bright quasars exhibiting blue-shifted absorption lines,
which are interpreted as the absorption by
the outflow material moving with velocity on the order of 
$0.1c$ (\cite{Pounds03}; \cite{Reeves03}). 
Although no strong outflows were expected
in the framework of the standard disk model,
\citet{Miller06} concluded by X-ray observations of BHB, 
GRO J1655-40,
that the X-ray absorbing wind is ejected from the geometrically thin disk
viewed at an inclination angle of $\sim 70^\circ$.
The density and the velocity of the wind were reported to be 
$\sim 10^{-9}{\rm g\,cm^{-3}}$ and $0.001c-0.05c$.
Such features are roughly consistent with our simulation (model B),
in which the gas with 
$\rho={\rm several}\times 10^{-10}{\rm g\,cm^{-3}}$ 
is blown away towards the diagonal direction 
(the polar angle of $\sim 45^\circ$)
at the speed of $\sim 0.01c$. 
Our simulations show that the magnetic field lines 
are along the jet axis (models A and C).
Such structure is revealed by the 
recent radio observations of 
polarized emission in BL Lac object, Markarian 501
\citep{Giroletti08}.
The detailed comparison with the observations is left as future work.

To summarize, 
our global RMHD simulations can open a new era of accretion disk research 
and provide a unified view of accretion flows in various contexts.

\bigskip
\bigskip
The computations were performed on 
a parallel computer at Rikkyo University, 
XT4 system at CfCA of NAOJ, 
Super Combined Cluster System at RIKEN, 
and the computational facilities including the T2K Tsukuba 
and FIRST cluster at CCS in University of Tsukuba. 
This work is supported in part 
by Ministry of Education, Culture, Sports, Science, and 
Technology (MEXT) Young Scientist (B) 20740115 (KO), 
by special postdoctoral researchers program in RIKEN (KO), 
by the Grant-in-Aid of MEXT (19340044, SM), 
by the Grant-in-Aid for the global COE programs on 
"The Next Generation of Physics, Spun from Diversity and Emergence" 
from MEXT (SM)
and by the Grant-in-Aid of JSPS (14740132, MM).

\end{document}